\magnification=\magstep1


\newbox\SlashedBox 
\def\slashed#1{\setbox\SlashedBox=\hbox{#1}
\hbox to 0pt{\hbox to 1\wd\SlashedBox{\hfil/\hfil}\hss}{#1}}
\def\hboxtosizeof#1#2{\setbox\SlashedBox=\hbox{#1}
\hbox to 1\wd\SlashedBox{#2}}

\def\mathslashed#1{\setbox\SlashedBox=\hbox{$#1$}
\hbox to 0pt{\hbox to 1\wd\SlashedBox{\hfil/\hfil}\hss}#1}

\def\ifsmall{\iffalse}  
\def\titlepagefont{}  

\def\DefineTeXgraphics{%
\special{ps::[global] /TeXgraphics { } def}}  

\def\today{\ifcase\month\or January\or February\or March\or April\or May
\or June\or July\or August\or September\or October\or November\or
December\fi\space\number\day, \number\year}
\def\eatPrefix19{}
\def\Year{\expandafter\eatPrefix\the\year}
\newcount\hours \newcount\minutes
\def\monthname{\ifcase\month\or
January\or February\or March\or April\or May\or June\or July\or
August\or September\or October\or November\or December\fi}
\def\shortmonthname{\ifcase\month\or
Jan\or Feb\or Mar\or Apr\or May\or Jun\or Jul\or
Aug\or Sep\or Oct\or Nov\or Dec\fi}

\def\TimeStamp{\hours\the\time\divide\hours by60%
\minutes -\the\time\divide\minutes by60\multiply\minutes by60%
\advance\minutes by\the\time%
${\rm \shortmonthname}\cdot\if\day<10{}0\fi\the\day\cdot\the\year%
\qquad\the\hours:\if\minutes<10{}0\fi\the\minutes$}




 


\newif\ifdraftmode
\newif\ifleftlabels  

\def\nolabels{\def\wrlabeL##1{}\def\eqlabeL##1{}\def\reflabeL##1{}}
\def\writelabels{\def\wrlabeL##1{\leavevmode\vadjust{\rlap{\smash%
{\line{{\escapechar=` \hfill\rlap{\sevenrm\hskip.03in\string##1}}}}}}}%
\def\eqlabeL##1{{\escapechar-1\rlap{\sevenrm\hskip.05in\string##1}}}%
\def\reflabeL##1{\noexpand\rlap{\noexpand\sevenrm[\string##1]}}}
\def\writeleftlabels{\def\wrlabeL##1{\leavevmode\vadjust{\rlap{\smash%
{\line{{\escapechar=` \hfill\rlap{\sevenrm\hskip.03in\string##1}}}}}}}%
\def\eqlabeL##1{{\escapechar-1%
\rlap{\sixrm\hskip.05in\string##1}%
\llap{\sevenrm\string##1\hskip.03in\hbox to \hsize{}}}}%
\def\reflabeL##1{\noexpand\rlap{\noexpand\sevenrm[\string##1]}}}
\nolabels

\newdimen\fullhsize
\newdimen\hstitle
\hstitle=\hsize 
\newdimen\hsbody
\hsbody=\hsize 
\newdimen\hbodyoffset
\hbodyoffset=\hoffset 
\newbox\leftpage
\def\abstract#1{#1}
\def\rotated{\special{ps: landscape}
\magnification=1000  
\baselineskip=14pt
\global\hstitle=9truein\global\hsbody=4.75truein
\global\vsize=7truein\global\voffset=-.31truein
\global\hoffset=-0.54in\global\hbodyoffset=-.54truein
\global\fullhsize=10truein
\def\DefineTeXgraphics{%
\special{ps::[global] 
/TeXgraphics {currentpoint translate 0.7 0.7 scale
              -80 0.72 mul -1000 0.72 mul translate} def}}
\let\lr=L
\def\ifsmall{\iftrue}
\def\titlepagefont{\twelvepoint}
\trueseventeenpoint
\def\almostshipout##1{\if L\lr \count1=1
      \global\setbox\leftpage=##1 \global\let\lr=R
   \else \count1=2
      \shipout\vbox{\hbox to\fullhsize{\box\leftpage\hfil##1}}
      \global\let\lr=L\fi}

\output={\ifnum\count0=1 
 \shipout\vbox{\hbox to \fullhsize{\hfill\pagebody\hfill}}\advancepageno
 \else
 \almostshipout{\leftline{\vbox{\pagebody\makefootline}}}\advancepageno 
 \fi}

\def\abstract##1{{\leftskip=1.5in\rightskip=1.5in ##1\par}} }

\def\linemessage#1{\immediate\write16{#1}}

\global\newcount\secno \global\secno=0
\global\newcount\appno \global\appno=0
\global\newcount\meqno \global\meqno=1
\global\newcount\subsecno \global\subsecno=0
\global\newcount\figno \global\figno=0

\newif\ifAnyCounterChanged
\let\terminator=\relax
\def\normalize#1{\ifx#1\terminator\let\next=\relax\else%
\if#1i\aftergroup i\else\if#1v\aftergroup v\else\if#1x\aftergroup x%
\else\if#1l\aftergroup l\else\if#1c\aftergroup c\else%
\if#1m\aftergroup m\else%
\if#1I\aftergroup I\else\if#1V\aftergroup V\else\if#1X\aftergroup X%
\else\if#1L\aftergroup L\else\if#1C\aftergroup C\else%
\if#1M\aftergroup M\else\aftergroup#1\fi\fi\fi\fi\fi\fi\fi\fi\fi\fi\fi\fi%
\let\next=\normalize\fi%
\next}
\def\makeNormal#1#2{\def\doNormalDef{\edef#1}\begingroup%
\aftergroup\doNormalDef\aftergroup{\normalize#2\terminator\aftergroup}%
\endgroup}

\def\warnIfChanged#1#2{%
\ifundef#1
\else\begingroup%
\edef\oldDefinitionOfCounter{#1}\edef\newDefinitionOfCounter{#2}%
\ifx\oldDefinitionOfCounter\newDefinitionOfCounter%
\else%
\linemessage{Warning: definition of \noexpand#1 has changed.}%
\global\AnyCounterChangedtrue\fi\endgroup\fi}

\def\Section#1{\global\advance\secno by1\relax\global\meqno=1%
\global\subsecno=0%
\bigbreak\bigskip
\centerline{\twelvepoint \bf %
\the\secno. #1}%
\par\nobreak\medskip\nobreak}
\def\tagsection#1{%
\warnIfChanged#1{\the\secno}%
\xdef#1{\the\secno}%
\ifWritingAuxFile\immediate\write\auxfile{\noexpand\xdef\noexpand#1{#1}}\fi%
}
\def\section{\Section}
\def\Subsection#1{\global\advance\subsecno by1\relax\medskip %
\leftline{\bf\the\secno.\the\subsecno\ #1}%
\par\nobreak\smallskip\nobreak}
\def\tagsubsection#1{%
\warnIfChanged#1{\the\secno.\the\subsecno}%
\xdef#1{\the\secno.\the\subsecno}%
\ifWritingAuxFile\immediate\write\auxfile{\noexpand\xdef\noexpand#1{#1}}\fi%
}

\def\subsection{\Subsection}

\def\romappno{\uppercase\expandafter{\romannumeral\appno}}
\def\makeNormalizedRomappno{%
\expandafter\makeNormal\expandafter\normalizedromappno%
\expandafter{\romannumeral\appno}%
\edef\normalizedromappno{\uppercase{\normalizedromappno}}}
\def\Appendix#1{\global\advance\appno by1\relax\global\meqno=1\global\secno=0%
\global\subsecno=0%
\bigbreak\bigskip
\centerline{\twelvepoint \bf Appendix %
\romappno. #1}%
\par\nobreak\medskip\nobreak}
\def\tagappendix#1{\makeNormalizedRomappno%
\warnIfChanged#1{\normalizedromappno}%
\xdef#1{\normalizedromappno}%
\ifWritingAuxFile\immediate\write\auxfile{\noexpand\xdef\noexpand#1{#1}}\fi%
}
\def\appendix{\Appendix}
\def\Subappendix#1{\global\advance\subsecno by1\relax\medskip %
\leftline{\bf\romappno.\the\subsecno\ #1}%
\par\nobreak\smallskip\nobreak}
\def\tagsubappendix#1{\makeNormalizedRomappno%
\warnIfChanged#1{\normalizedromappno.\the\subsecno}%
\xdef#1{\normalizedromappno.\the\subsecno}%
\ifWritingAuxFile\immediate\write\auxfile{\noexpand\xdef\noexpand#1{#1}}\fi%
}

\def\eqn#1{\makeNormalizedRomappno%
\ifnum\secno>0%
  \warnIfChanged#1{\the\secno.\the\meqno}%
  \eqno(\the\secno.\the\meqno)\xdef#1{\the\secno.\the\meqno}%
     \global\advance\meqno by1
\else\ifnum\appno>0%
  \warnIfChanged#1{\normalizedromappno.\the\meqno}%
  \eqno({\rm\romappno}.\the\meqno)%
      \xdef#1{\normalizedromappno.\the\meqno}%
     \global\advance\meqno by1
\else%
  \warnIfChanged#1{\the\meqno}%
  \eqno(\the\meqno)\xdef#1{\the\meqno}%
     \global\advance\meqno by1
\fi\fi%
\eqlabeL#1%
\ifWritingAuxFile\immediate\write\auxfile{\noexpand\xdef\noexpand#1{#1}}\fi%
}
\def\defeqn#1{\makeNormalizedRomappno%
\ifnum\secno>0%
  \warnIfChanged#1{\the\secno.\the\meqno}%
  \xdef#1{\the\secno.\the\meqno}%
     \global\advance\meqno by1
\else\ifnum\appno>0%
  \warnIfChanged#1{\normalizedromappno.\the\meqno}%
  \xdef#1{\normalizedromappno.\the\meqno}%
     \global\advance\meqno by1
\else%
  \warnIfChanged#1{\the\meqno}%
  \xdef#1{\the\meqno}%
     \global\advance\meqno by1
\fi\fi%
\eqlabeL#1%
\ifWritingAuxFile\immediate\write\auxfile{\noexpand\xdef\noexpand#1{#1}}\fi%
}
\def\anoneqn{\makeNormalizedRomappno%
\ifnum\secno>0
  \eqno(\the\secno.\the\meqno)%
     \global\advance\meqno by1
\else\ifnum\appno>0
  \eqno({\rm\normalizedromappno}.\the\meqno)%
     \global\advance\meqno by1
\else
  \eqno(\the\meqno)%
     \global\advance\meqno by1
\fi\fi%
}
\def\mfig#1#2{\ifx#20
\else\global\advance\figno by1%
\relax#1\the\figno%
\warnIfChanged#2{\the\figno}%
\xdef#2{\the\figno}%
\reflabeL#2%
\ifWritingAuxFile\immediate\write\auxfile{\noexpand\xdef\noexpand#2{#2}}\fi\fi%
}

\catcode`@=11 

\newif\ifFiguresInText\FiguresInTexttrue
\newif\if@FigureFileCreated
\newwrite\capfile
\newwrite\figfile

\newif\ifcaption
\captiontrue
\def\captionsize{\tenrm}
\def\PlaceTextFigure#1#2#3#4{%
\vskip 0.5truein%
#3\hfil\epsfbox{#4}\hfil\break%
\ifcaption\hfil\vbox{\captionsize Figure #1. #2}\hfil\fi%
\vskip10pt}
\def\PlaceEndFigure#1#2{%
\epsfxsize=\hsize\epsfbox{#2}\vfill\centerline{Figure #1.}\eject}

\def\LoadFigure#1#2#3#4{%
\ifundef#1{\phantom{\mfig{}#1}}\else
\fi%
\ifFiguresInText
\PlaceTextFigure{#1}{#2}{#3}{#4}%
\else
\if@FigureFileCreated\else%
\immediate\openout\capfile=\jobname.caps%
\immediate\openout\figfile=\jobname.figs%
@FigureFileCreatedtrue\fi%
\immediate\write\capfile{\noexpand\item{Figure \noexpand#1.\ }{#2}\vskip10pt}%
\immediate\write\figfile{\noexpand\PlaceEndFigure\noexpand#1{\noexpand#4}}%
\fi}

\def\listfigs{\ifFiguresInText\else%
\vfill\eject\immediate\closeout\capfile
\immediate\closeout\figfile%
\centerline{{\bf Figures}}\bigskip\frenchspacing%
\catcode`@=11 
\def\captionsize{\tenrm}
\input \jobname.caps\vfill\eject\nonfrenchspacing%
\catcode`\@=\active
\catcode`@=12  
\input\jobname.figs\fi}

\font\ninerm=cmr9
\font\eightrm=cmr8
\font\sixrm=cmr6

\def\loadtrueseventeenpoint{
 \font\seventeenrm=cmr10 at 17.28truept
 \font\seventeeni=cmmi10 at 17.28truept
 \font\seventeenbf=cmbx10 at 17.28truept
 \font\seventeenit=cmti10 at 17.28truept
 \font\seventeensl=cmsl10 at 17.28truept
 \font\seventeensy=cmsy10 at 17.28truept
}
\def\loadfourteenpoint{
\font\fourteenrm=cmr10 at 14.4pt
\font\fourteeni=cmmi10 at 14.4pt
\font\fourteenit=cmti10 at 14.4pt
\font\fourteensl=cmsl10 at 14.4pt
\font\fourteensy=cmsy10 at 14.4pt
\font\fourteenbf=cmbx10 at 14.4pt
}
\def\loadtruetwelvepoint{
\font\twelverm=cmr10 at 12truept
\font\twelvei=cmmi10 at 12truept
\font\twelveit=cmti10 at 12truept
\font\twelvesl=cmsl10 at 12truept
\font\twelvesy=cmsy10 at 12truept
\font\twelvebf=cmbx10 at 12truept
}

\font\ninei=cmmi9
\font\eighti=cmmi8
\font\sixi=cmmi6
\skewchar\ninei='177 \skewchar\eighti='177 \skewchar\sixi='177

\font\ninesy=cmsy9
\font\eightsy=cmsy8
\font\sixsy=cmsy6
\skewchar\ninesy='60 \skewchar\eightsy='60 \skewchar\sixsy='60

\font\ninebf=cmbx9
\font\eightbf=cmbx8
\font\sixbf=cmbx6

\font\ninett=cmtt9
\font\eighttt=cmtt8

\hyphenchar\tentt=-1 
\hyphenchar\ninett=-1
\hyphenchar\eighttt=-1         

\font\ninesl=cmsl9
\font\eightsl=cmsl8

\font\nineit=cmti9
\font\eightit=cmti8

                      
\newskip\ttglue
\def\tenpoint{\def\rm{\fam0\tenrm}%
  \textfont0=\tenrm \scriptfont0=\sevenrm \scriptscriptfont0=\fiverm
  \textfont1=\teni \scriptfont1=\seveni \scriptscriptfont1=\fivei
  \textfont2=\tensy \scriptfont2=\sevensy \scriptscriptfont2=\fivesy
  \textfont3=\tenex \scriptfont3=\tenex \scriptscriptfont3=\tenex
  \def\it{\fam\itfam\tenit}\textfont\itfam=\tenit
  \def\sl{\fam\slfam\tensl}\textfont\slfam=\tensl
  \def\bf{\fam\bffam\tenbf}\textfont\bffam=\tenbf \scriptfont\bffam=\sevenbf
  \scriptscriptfont\bffam=\fivebf
  \normalbaselineskip=12pt
  \let\sc=\eightrm
  \let\big=\tenbig
  \setbox\strutbox=\hbox{\vrule height8.5pt depth3.5pt width\z@}%
  \normalbaselines\rm}

\def\twelvepoint{\def\rm{\fam0\twelverm}%
  \textfont0=\twelverm \scriptfont0=\ninerm \scriptscriptfont0=\sevenrm
  \textfont1=\twelvei \scriptfont1=\ninei \scriptscriptfont1=\seveni
  \textfont2=\twelvesy \scriptfont2=\ninesy \scriptscriptfont2=\sevensy
  \textfont3=\tenex \scriptfont3=\tenex \scriptscriptfont3=\tenex
  \def\it{\fam\itfam\twelveit}\textfont\itfam=\twelveit
  \def\sl{\fam\slfam\twelvesl}\textfont\slfam=\twelvesl
  \def\bf{\fam\bffam\twelvebf}\textfont\bffam=\twelvebf%
  \scriptfont\bffam=\ninebf
  \scriptscriptfont\bffam=\sevenbf
  \normalbaselineskip=12pt
  \let\sc=\eightrm
  \let\big=\tenbig
  \setbox\strutbox=\hbox{\vrule height8.5pt depth3.5pt width\z@}%
  \normalbaselines\rm}

\def\seventeenpoint{\def\rm{\fam0\seventeenrm}%
  \textfont0=\seventeenrm \scriptfont0=\fourteenrm \scriptscriptfont0=\tenrm
  \textfont1=\seventeeni \scriptfont1=\fourteeni \scriptscriptfont1=\teni
  \textfont2=\seventeensy \scriptfont2=\fourteensy \scriptscriptfont2=\tensy
  \textfont3=\tenex \scriptfont3=\tenex \scriptscriptfont3=\tenex
  \def\it{\fam\itfam\seventeenit}\textfont\itfam=\seventeenit
  \def\sl{\fam\slfam\seventeensl}\textfont\slfam=\seventeensl
  \def\bf{\fam\bffam\seventeenbf}\textfont\bffam=\seventeenbf%
  \scriptfont\bffam=\fourteenbf
  \scriptscriptfont\bffam=\twelvebf
  \normalbaselineskip=21pt
  \let\sc=\fourteenrm
  \let\big=\tenbig                                          
  \setbox\strutbox=\hbox{\vrule height 12pt depth 6pt width\z@}%
  \normalbaselines\rm}

\def\ninepoint{\def\rm{\fam0\ninerm}%
  \textfont0=\ninerm \scriptfont0=\sixrm \scriptscriptfont0=\fiverm
  \textfont1=\ninei \scriptfont1=\sixi \scriptscriptfont1=\fivei
  \textfont2=\ninesy \scriptfont2=\sixsy \scriptscriptfont2=\fivesy
  \textfont3=\tenex \scriptfont3=\tenex \scriptscriptfont3=\tenex
  \def\it{\fam\itfam\nineit}\textfont\itfam=\nineit
  \def\sl{\fam\slfam\ninesl}\textfont\slfam=\ninesl
  \def\bf{\fam\bffam\ninebf}\textfont\bffam=\ninebf \scriptfont\bffam=\sixbf
  \scriptscriptfont\bffam=\fivebf
  \normalbaselineskip=11pt
  \let\sc=\sevenrm
  \let\big=\ninebig
  \setbox\strutbox=\hbox{\vrule height8pt depth3pt width\z@}%
  \normalbaselines\rm}

\def\eightpoint{\def\rm{\fam0\eightrm}%
  \textfont0=\eightrm \scriptfont0=\sixrm \scriptscriptfont0=\fiverm%
  \textfont1=\eighti \scriptfont1=\sixi \scriptscriptfont1=\fivei%
  \textfont2=\eightsy \scriptfont2=\sixsy \scriptscriptfont2=\fivesy%
  \textfont3=\tenex \scriptfont3=\tenex \scriptscriptfont3=\tenex%
  \def\it{\fam\itfam\eightit}\textfont\itfam=\eightit%
  \def\sl{\fam\slfam\eightsl}\textfont\slfam=\eightsl%
  \def\bf{\fam\bffam\eightbf}\textfont\bffam=\eightbf \scriptfont\bffam=\sixbf%
  \scriptscriptfont\bffam=\fivebf%
  \normalbaselineskip=9pt%
  \let\sc=\sixrm%
  \let\big=\eightbig%
  \setbox\strutbox=\hbox{\vrule height7pt depth2pt width\z@}%
  \normalbaselines\rm}

\def\tenbig#1{{\hbox{$\left#1\vbox to8.5pt{}\right.\n@space$}}}
\def\ninebig#1{{\hbox{$\textfont0=\tenrm\textfont2=\tensy
  \left#1\vbox to7.25pt{}\right.\n@space$}}}
\def\eightbig#1{{\hbox{$\textfont0=\ninerm\textfont2=\ninesy
  \left#1\vbox to6.5pt{}\right.\n@space$}}}

\def\fourteenpoint{\def\rm{\fam0\fourteenrm}%
  \textfont0=\fourteenrm \scriptfont0=\tenrm \scriptscriptfont0=\sevenrm
  \textfont1=\fourteeni \scriptfont1=\teni \scriptscriptfont1=\seveni
  \textfont2=\fourteensy \scriptfont2=\tensy \scriptscriptfont2=\sevensy
  \textfont3=\tenex \scriptfont3=\tenex \scriptscriptfont3=\tenex
  \def\it{\fam\itfam\fourteenit}\textfont\itfam=\fourteenit
  \def\sl{\fam\slfam\fourteensl}\textfont\slfam=\fourteensl
  \def\bf{\fam\bffam\fourteenbf}\textfont\bffam=\fourteenbf%
  \scriptfont\bffam=\tenbf
  \scriptscriptfont\bffam=\sevenbf
  \normalbaselineskip=17pt
  \let\sc=\elevenrm
  \let\big=\tenbig                                          
  \setbox\strutbox=\hbox{\vrule height8.5pt depth3.5pt width\z@}%
  \normalbaselines\rm}

\def\footnote#1{\edef\@sf{\spacefactor\the\spacefactor}#1\@sf
      \insert\footins\bgroup\eightpoint
      \interlinepenalty100 \let\par=\endgraf
        \leftskip=\z@skip \rightskip=\z@skip
        \splittopskip=10pt plus 1pt minus 1pt \floatingpenalty=20000
        \smallskip\item{#1}\bgroup\strut\aftergroup\@foot\let\next}
\skip\footins=12pt plus 2pt minus 4pt 
\dimen\footins=30pc 

\newinsert\margin
\dimen\margin=\maxdimen

\loadtruetwelvepoint 
\loadtrueseventeenpoint

\def\eatOne#1{}
\def\ifundef#1{\expandafter\ifx%
\csname\expandafter\eatOne\string#1\endcsname\relax}
\def\notTrue{\iffalse}\def\isTrue{\iftrue}
\def\ifdef#1{{\ifundef#1%
\aftergroup\notTrue\else\aftergroup\isTrue\fi}}
\def\use#1{\ifundef#1\linemessage{Warning: \string#1 is undefined.}%
{\tt \string#1}\else#1\fi}



%
\catcode`"=11
\let\quote="
\catcode`"=12
\chardef\foo="22
\global\newcount\refno \global\refno=1
\newwrite\rfile
\newlinechar=`\^^J
\def\@ref#1#2{\the\refno\n@ref#1{#2}}
\def\h@ref#1#2#3{\href{#3}{\the\refno}\n@ref#1{#2}}
\def\n@ref#1#2{\xdef#1{\the\refno}%
\ifnum\refno=1\immediate\openout\rfile=\jobname.refs\fi%
\immediate\write\rfile{\noexpand\item{[\noexpand#1]\ }#2.}%
\global\advance\refno by1}
\def\nref{\n@ref} 
\def\ref{\@ref}   
\def\hrref{\h@ref}
\def\lref#1#2{\the\refno\xdef#1{\the\refno}%
\ifnum\refno=1\immediate\openout\rfile=\jobname.refs\fi%
\immediate\write\rfile{\noexpand\item{[\noexpand#1]\ }#2\semi}%
\global\advance\refno by1}
\def\cref#1{\immediate\write\rfile{#1\semi}}

\def\preref#1#2{\gdef#1{\@ref#1{#2}}}

\def\semi{;\hfil\noexpand\break}

\def\listrefs{\vfill\eject\immediate\closeout\rfile
\centerline{{\bf References}}\bigskip\frenchspacing%
\input \jobname.refs\vfill\eject\nonfrenchspacing}

\def\inputAuxIfPresent#1{\immediate\openin1=#1
\ifeof1\message{No file \auxfileName; I'll create one.
}\else\closein1\relax\input\auxfileName\fi%
}




\newif\ifWritingAuxFile
\newwrite\auxfile
\def\SetUpAuxFile{%
\xdef\auxfileName{\jobname.aux}%
\inputAuxIfPresent{\auxfileName}%
\WritingAuxFiletrue%
\immediate\openout\auxfile=\auxfileName}



\catcode`\@=\active
\catcode`@=12  
\catcode`\"=\active


\SetUpAuxFile 

\preref\dpone{
E.\ D'Hoker, D.H.\ Phong, Nucl.\ Phys.\ B440:24 (1995), hep-th/9410152} 

\preref\dptwo{
E.\ D'Hoker, D.H.\ Phong, Theor.\ Math.\ Phys.\ 98:306 (1994), hep-th/9404128}

\preref\dpthree{
E.\ D'Hoker, D.H.\ Phong, Phys.\ Rev.\ Lett. 70:3692 (1993), hep-th/9302003}  

\preref\gsb{
M.\ Green, J.H.\ Schwarz, L.\ Brink, Nucl.\ Phys.\ B198:474 (1982)} 




\preref\gsw{ 
M.\ Green, J.H.\ Schwarz, E.\ Witten, Superstring Theory, (1986)}  

\preref\bklong{ 
Z.\ Bern, D.\ Kosower, Nucl.\ Phys.\ B379:451 (1992)} 

\preref\kk{  
E.\ Kiritsis, C.\ Kounnas Nucl.\ Phys.\ B442:472 (1995), hep-th/9501020}

\preref\dimreg{ 
G.\ 't Hooft, M.\ Veltman, Nucl.\ Phys.\ B44:189 (1972)}

\preref\dimred{ 
W.\ Siegel, Phys.\ Lett.\ 84B:193 (1979); Phys.\ Lett.\ 94B:37 (1980)}    

\preref\dprev{ 
E.\ D'Hoker, D.H.\ Phong, Rev.\ Mod.\ Phys.60:917 (1988)}  

\preref\bdkint{ 
Z.\ Bern, L.\ Dixon, D.\ Kosower, Nucl.\ Phys.\ B412:751 (1994), hep-ph/9306240}

\preref\ars{ 
J.\ Atick, J.\ Rabin, A.\ Sen, Nucl.\ Phys.\ B299:279 (1987)}   

\preref\cdp{ 
G.\ Chalmers, E.\ D'Hoker, D.\ Phong, unpublished} 

\preref\tata{  
D.\ Mumford, Tata Lecture Notes on Theta Functions, Birkhauser, 
Cambridge (1983)}  

\preref\parttwo{ 
G.\ Chalmers, in progress}  

\preref\gs{ 
M.B.\ Green, J.H.\ Schwarz, Phys.\ Lett.\ B109:444 (1982)}



\preref\mw{ 
J.L.\ Montag, W.\ Weisberger, Nucl.\ Phys.\ B363:527 (1991)}  

\preref\polch{ 
J.\ Polchinski, Comm.\ Math.\ Phys.\ 104:37 (1986)} 

\preref\dpfour{ 
E.\ D'Hoker, D.H.\ Phong, Nucl.\ Phys.\ B278:225 (1986)} 

\preref\mnp{ 
G.\ Moore, P.\ Nelson, J.\ Polchinski, Phys.\ Lett.\ B169:47 (1986)}


\baselineskip=14pt

\def\pol{\varepsilon}
\def\eps{\epsilon}
\def\c{\cdot}
\def\half{{1\over 2}}

\rightline{ITP-SB-97-74}
\rightline{hep-th yymmdd}

\vglue .5cm
\centerline{{\bf Momumentum Analyticity and Finiteness of}} 
\centerline{{\bf Compactified String Amplitudes}} 
\vskip .05in
\centerline{{\bf Part I: Tori}}

\vglue 1cm
\centerline{{\bf Gordon Chalmers}} \vskip .2cm
\baselineskip=13pt
\centerline{\it Institute for Theoretical Physics} 
\centerline{\it State University of New York} 
\centerline{\it Stony Brook, NY   11794-3840}  
\centerline{\it chalmers@insti.physics.sunysb.edu} 

\vglue 2cm

\centerline{\bf ABSTRACT}
\vglue 0.3cm

We generalize to the case of compactified superstrings a  construction given previously for
critical superstrings of finite one loop amplitudes that are well-defined for all external
momenta. The novel issues that arise for compactified strings are the appearance of infrared
divergences from the propagation of massless strings in four dimensions and, in the  case of
orbifold schemes, the contribution of tachyons in partial amplitudes with given spin structure
and twist sectors. Methods are presented for the resolution of both problems and expressions
for finite amplitudes are given in terms of double and single dispersion relations, with
explicit spectral densities. 

\vfil  

\baselineskip=12pt 
\loadfourteenpoint  
\hfuzz=50pt

\noindent  
\section{Introduction} 
\vskip .3in

In previous papers [\use\dpone,\use\dptwo,\use\dpthree ], it was pointed out that one-loop 
four-point amplitudes in heterotic or Type II superstrings in their critical dimension are
properly defined by the standard integral representation over moduli only for purely imaginary
values of the Mandelstam variables $s_{ij}$ [\use\dpone,\use\mw ]. Away from imaginary $s_{ij}$, the
integral representation is divergent and, for real (physical) values of $s_{ij}$, it is formally real.
Both the reality and the divergence of the one-loop amplitude are physically unacceptable. The
imaginary part of the (forward) one loop amplitude is related -- by unitarity -- to the
absolute value squared of the tree level four point function, which is of course known to be
non-zero.

Both problems were solved in [1] for the critical string, by showing that there  exists an
analytic continuation of the integral representation to all $s_{ij}$ which exhibits precisely
the expected poles and branch cuts in $s_{ij}$. An explicit construction of this analytic
continuation was given in [\use\dpone ] for the case of the four-point function 
of states in the gravitational multiplet.

In the present paper, we shall extend the analysis used to deal with these 
problems to the case of
superstrings propagating on four dimensional Minkowski space-time
$M_4$ times a six dimensional compactified space $K_6$. A number of important novel features
arise for compactified strings that did not present themselves for the critical superstring.

First, the presence of massless particles in four  dimensions generically produces infrared
(IR) divergences in the scattering amplitudes. IR divergences are familiar already from QED,
where they signal the fact that physical charged states contain admixtures of soft photons.
The Bloch-Nordsieck theorem guarantess that these IR divergences cancel out of the calculation
of any physical cross section. In practice, an IR regulator may be introduced which produces
finite but regulator dependent amplitudes. Physical cross sections should be finite once all
the relevant contrinutions have been taken into account and they should be insensitive to the
specific regulator used. If the theory contains charged massless particles, as in unbroken
non-Abelian gauge theories, the IR divergences become even more severe, and great care must be
used to make the IR regulator consistent with all symmetries. For gauge theories, dimensional
regularization is in general a suitable regulator, while in supersymmetric theories,
dimensional reduction is preferred. 

Amplitudes derived from compactified superstring theories are also generally IR divergent,
since they contain an unbroken non-Abelian gauge sector as well as gravity. Compared to the
critical superstring, the IR divergences occurring for compactified strings present an extra
source of infinities (independent of $s_{ij}$), not yet encountered in the analysis of
[\use\dpone,\use\dptwo, \use\dpthree ]. It is a rather subtle issue to obtain a consistent IR
regularization scheme in compactified superstring theory, where modular invariance is of
crucial importance. Two regularization schemes that have already been discussed in the
literature will be applied here, but both have different drawbacks that we shall point out. We
shall show in subsequent work that both schemes lead to well-defined, finite analytic
continuations of the compactified amplitudes, consistent with modular invariance.

Second, in the case of orbifold compactifications, extra divergences may arise from the  presence of
the tachyon in certain partial amplitudes that only involve the Neveu-Schwarz (NS) sector, or that
involve only one twisting. Such tachyon related divergences
in fact already arise in the uncompactified critical superstring amplitudes, when partial amplitudes
of definite spin structure are considered.  For the four-point ampitludes, where
spin structure sums are easily carried out explicitly, these tachyon divergences did not have
to be dealt with explicitly.  Also, for orbifold compactifications, the way the tachyon enters
is more complicated than in the uncompactified superstring.  By a suitable
grouping together of expansion terms of the amplitudes, these tachyon pole and branch cut
contributions are easily isolated and neutralized.  

In this work we shall consider only the toroidal compactifications of either 
the IIb superstring or the Heterotic model and produce the analagous results 
for the case of orbifold models in a different study.


\vskip .2in
\section{Amplitudes in Toroidal Compactifications} 
\medskip

We shall consider only
amplitudes for external states that have remained massless after the compactification:  In 
the toroidal case, these states include the graviton, a number of Kaluza-Klein $U(1)$ gauge
fields that arise from compactifying the metric $G_{\mu i}$ or the anti-symmetric tensor field
$B_{\mu i}$ (including the axion).  All these states arise from NS-NS states in the
uncompactified string and are represented by vertex operators in which the external momenta
for the compactified directions are set to zero. They are particularly simple and the
calculation of amplitudes involving them is simplified by the fact that the dependence upon
the external vertex operators is identical to that of the critical superstring, with
non-trivial toroidal states propagating only inside the loop.  Restriction to these 
states thus only requires further evaluation of the partition function for the zero 
modes of the strings, to be inserted into the integrand for the critical superstring.  

The partition function (on a torus worldsheet with  complex modulus $\tau$) evaluated for
strings compactified on a torus, were given for both the Type II and heterotic strings as a
sum over the momenta belonging to the lattice $\Gamma$ that characterizes the
compactification : 
$$
\eqalign{ Z(\Gamma) = & ({\rm Im}~ \tau )^{{10-d}\over 2} {\tilde Z}(\Gamma) 
\cr {\tilde Z} (\Gamma) = & \sum_{(P_L,P_R)\in \Gamma} e^{i\pi\tau P_{L} \c P_{L} - i\pi \bar
\tau P_{R}\c P_{R} } \cr }  \ .
\eqn\lattice
$$  In (\use\lattice), the momenta $(P_L,P_R)$ parameterize the Lorentzian lattice $\Gamma$ 
with signature $(p,q)$. Modular invariance of $Z(\Gamma)$ requires
$\Gamma$ to be self-dual and even, in the sense that the quadratic form $(P_L^2 -P_R ^2) /2$ is
integer valued.  Such lattices are known to exist only when $p-q$ is an integer multiple of 8
[\use\gsw ].

For the Type II string compactified on a symmetric lattice, we have $p=q=10-d$, which certainly
satisfies this condition; furthermore, all lattices $\Gamma$ are obtained by taking $p$ copies
of the two-dimensional lattice $\Gamma _2$ with signature $(1,1)$.  The geometric means of
defining the lattice is through the background flat metric 
$g_{ij}$ and constant anti-symmetric tensor field $B_{ij}$ on the $10-d$-dimensional  torus: 
The lattice momenta are conveniently parametrized in terms of these as follows :
$$
\eqalign{ P_L ^i =& n^i +\half g^{ij} m_j - g^{ij} B_{jk} n^k 
\cr P_R ^i =& n^i -\half g^{ij} m_j + g^{ij} B_{jk} n^k \cr } \ ,
\eqn\latticemom 
$$  where $g^{ij}$ is the inverse matrix of $g_{ij}$ with $n^i$  and $m_j$ integers. (The
metric
$g_{ij}$ may be viewed as the  matrix of inner products of the basis vectors of the lattice
$\Gamma$.)

In the case of the heterotic string, one has $p=26-d$, $q=10-d$, and  for $q>0$, all lattices
are obtained as sums of two copies of the root lattice $\Gamma _8$ of $E_8$ and $q$ copies of
the two-dimensional lattice $\Gamma _2$ introduced above : $\Gamma =2\Gamma _8\oplus q\Gamma
_2$. (For the uncompactified heterotic string with
$q=0$, there is also the lattice $\Gamma _{16}$ associated with $Spin(32)/Z_2$.)  The lattice
momenta may now be parametrized in terms of the metric $g_{ij}$ on the $10-d$ dimensional
torus, the anti-symmetric tensor field $B_{ij}$ and the 16 Abelian gauge fields $A_i ^I$ which
belong to the Cartan subalgebra of either $Spin(32)/Z_2$ or $E_8\times E_8$ and arise from
commuting Wilson lines on the torus.
$$
\eqalign{ P_L ^i =& n^i +\half g^{ij} m_j - g^{ij} B_{jk} n^k + 
\cdots \cr P_R ^i =& n^i -\half g^{ij} m_j + g^{ij} B_{jk} n^k + \cdots \cr P_R ^I =&
\cdots \cr } 
\eqn\heterlatticemom 
$$  With the help of these parametrizations, the partition function $Z(\Gamma)$ can be
evaluated in terms of a convergent sum over powers $q^{r} \bar q ^{\bar r}$ (where as usual 
$q=e^{2i\pi\tau}$) as was done for the simple case of the toroidal compactification.  

One-loop amplitudes in the toroidally compactified theory with $N$ external massless vertex
operators (for states in the NS-NS sector), and evaluated at vanishing compactification
momentum, are obtained simply by substituting the lattice partition function
$Z(\Gamma)$ into the integrals for the corresponding critical superstring amplitudes.  We 
also must, of course,  restrict the momenta of the external states to the uncompactified
directions. Notice that the associated tree level amplitudes are unmodified for the massless
vertex operators we consider. We shall concentrate on the 4-point function for the scattering
of states  contained in the massless NS-NS multiplet [\use\gs,\use\polch,\use\dpfour, 
\use\mnp ], given as
follows for the Type II superstrings 
$$   A_{II} (k_i, \epsilon_i ) = (2\pi)^d \delta(k)g^4 A_{II} (s,t,u) K_{\mu_1\mu_2 \mu_3
\mu_4} K_{\bar \mu_1 \bar \mu_2 \bar \mu_3 \bar
\mu_4} \prod _{i=1} ^4 \epsilon_i ^{\mu_i\bar\mu_i} (k_i)  \ ,
\eqn\twobamp 
$$   where the factors $K$ depend on the particular states involved and $A_{II}(s,t,u)$  is a
universal scalar function depending only on the kinematics.  

For the heterotic string, we concentrate on the scattering amplitude for gauge bosons with root
(weight) lattice $K_i^I$, given by 
$$  A_{H} (k_i, \epsilon_i, K_i ) = (2\pi)^d
\delta(k)g^4 A_{H} (s,t,u;S,T,U) K_{\mu_1\mu_2 \mu_3 \mu_4}
\prod _{i=1} ^4 \epsilon_i ^{\mu_i} (k_i)  
\eqn\hetamp 
$$  Here, $g$ is the string coupling constant, $k=k_1+k_2+k_3+k_4$,  with $k_i ^\mu$ massless
momenta in the $d$ uncompactified dimensions, and external states characterized by the on-shell
conditions \footnote{*}{Repeated Lorentz indices will be assumed to be summed over throughout,
whereas repeated particle identification indices will not be assumed to be summed over.} 
$$   k_i ^\mu \epsilon _i ^\mu =0 \qquad
\qquad k_i^\mu \epsilon_i^{\mu N} (k_i) = k_i ^\nu \pol _i ^{M\nu}= 0
\qquad \qquad N=1,\cdots,10  
\anoneqn 
$$  The kinematical factor $K$ is a polynomial in momenta [\use\gsw ]  and $s=s_{12}=s_{34}$,
$t=s_{23}=s_{14}$, and $u=s_{13}=s_{24}$ are the familiar Mandelstam variables associated with
the momenta $k_i$ with $s+t+u = 0$, while for the heterotic string $S,~T,~U$ are internal root
(weight) lattice Mandelstam variables.

The Type {\rm IIb} one-loop amplitude for a compactification scheme with lattice $\Gamma$ of
signature
$(10-d,10-d)$ is given by ([\use\gs, \use\gsb, \use\gsw ]), 
$$  A_{II}(s,t,u) = \int_{F} {d^2\tau \over \tau_2^2} 
\int \prod^{4}_{i=1} {d^2z_i \over \tau _2} 
\exp \{ \half s_{ij} G(z_i,z_j) \} Z_{{\rm reg}}(\Gamma) \ .
\eqn\twobint 
$$  The corrresponding one loop amplitude for the heterotic string for states labelled by
$K_j$ is given by 
$$ A_{H}(s,t,u;S,T,U) = \int_{F} {d^2\tau \over \tau_2^2} 
\int \prod^{4}_{i=1} {d^2z_i \over \tau _2} 
\exp \{ \half s_{ij} G(z_i,z_j) \} \overline{Z_{\rm reg} (K_iz_i,\tau, \Gamma)} \ ,
\eqn\hetint 
$$  where $\Gamma$ is a lattice of signature $(10-d,26-d)$.  The integration region $F$ is the
fundamental moduli domain of the torus,
$$  F = \{ \tau= \tau_1 + i \tau_2 :~ \tau_2 > 0, ~~ -{1\over 2} \leq \tau_1 
\leq {1\over 2}, ~~ \vert \tau \vert \geq 1 \} \ ,
\anoneqn 
$$  and 
$$  G(z_i, z_j) = - \ln \biggl \vert {\vartheta_1 \bigl ( z_i - z_j, \tau \bigr) \over 
\vartheta'_1 \bigl( 0, \tau) } \biggr \vert^2 -  {\pi\over 2\tau_2} (z_i - {\bar z}_i - z_j +
{\bar z}_j )^2 \ ,  
\eqn\twopoint 
$$  is the scalar Green function on the torus.  The theta function $\vartheta_1(z,\tau)$ 
is defined for even characteristics [\use\tata ].  

The integral expressions for $A_{II}$ and $A_{H}$ contain infra-red divergences; one  sees
immediately from the fact that there are propagating massless internal states.    For this
reason we have not denoted the partition functions in the above as $Z(\Gamma)$, given in
eq.~(\use\lattice), but rather $Z_{{\rm reg}}(\Gamma)$, stressing the fact that a well-defined
amplitude is obtained only after suitable IR regularization. The corresponding 
regularized function
$Z_{\rm reg}(K_iz_i,\tau,\Gamma)$ contains the lattice sum required for the heterotic string,
given by
$$  {\cal L}(K_i,z_i,\tau,\Gamma) = \eta (\tau )^{-24} 
\prod _{i<j} \biggl ( {\vartheta _1 (z_i-z_j|\tau) \over \vartheta '_1(0|\tau) }\biggr ) ^{-K_i
\cdot K_j} \vartheta _\Gamma (K_iz_i,\tau)  
\anoneqn 
$$  The $\Gamma$ lattice
$\vartheta$-function is now not holomorphic in $\tau$ and given by
$$
\vartheta _\Gamma (K_iz_i|\tau) = \sum _{(P_L,P_R) \in \Gamma} 
\exp \bigl \{ i \pi \tau (P_L - {1 \over \tau} K_i z_i)^2 -i\pi 
\bar \tau P_R^2 \bigr \} \ ,
\anoneqn
$$   which arises from the $(p,q)=(26-d,10-d)$ dimensional lattice described  earlier.  

\vskip .2in
\section{IR Regularization}  
\vskip .2in

The amplitudes defined above through the formal integration over moduli  space are not well
defined due to the presence of infrared singularities and  must be regularized.   Infrared
regularization of string theory was originally applied in the study of the zero slope limit of
compactified theories [\use\gsb ]. It has also been widely used in string-inspired
simplifications of quantum field theory Feynman rules [\use\bklong ]. Most of these IR
regularization methods have been developed on a case by case basis, but no simple systematic
treatment of them or prescription for them seems to be available for  the string. We propose
that the minimal consistency requirements for a good regulator be as follows :

\vskip .2in
\item{(1)} The regulator must make all amplitudes finite and  well-defined with the help of
some regulator parameter, after suitable analytic continuation in the external momenta;
\item{(2)} It must correspond to a local action on the worldsheet and preserve modular
invariance;
\item{(3)} The naive amplitude must be recovered in the limit of vanishing regularization
parameter. 
\vskip .2in

Before we enter into the IR regularization scheme based on continuing in the dimensions of the
target torus, we shall briefly comment on another interesting IR regularization scheme that was
proposed in [\use\kk ]. This regulator satisfies criteria that are more
restrictive than the ones we have proposed here. In particular, it is required that the entire
string spectrum have a non-zero mass gap, which would certainly prohibit the 
entrance of infrared singularities. This criterion seems
too restrictive since, in the field theory limit, would exclude the use of ordinary
dimensional regularization for IR divergences. Also, the compactified string theory is
considered in non-trivial four-dimensional backgrounds, which leads to rather complicated conformal
field theories.  Finally, the proposal appeals to the use of amplitudes of non-critical string
theory that involve quantization of the Liouville field.  While much progress has been made on
this problem, it is certainly no straightforward task to calculate such Liouville amplitudes,
even in the limit of vanishing regulator. Thus, while certainly in the cases considered in
[\use\kk ] this regulator is consistent and has been applied successfully, in the cases of
toroidal and orbifold compactification that we consider here, it does not appear necessary to
make use of this complicated scheme.

Instead, for both the toroidal and orbifold compactification schemes, a slightly altered
version of dimensional regularization/dimensional reduction [\use\dimred, \use\dimreg ]
satisfies all the requirements we have proposed above and will ultimately lead to well-defined
scattering amplitudes. The method simply consists in analytically continuing the
uncompactified dimension from $d=4$ to $d=4+2\eps$ and the compactified dimensions from 6 to
$6-\eps$. The analytic continuation of the toroidally or orbifold compactified dimensions
poses no particular problem, as it may be formulated as the extension to fractional dimensions
of the theory on a torus with variable constant background metric and anti-symmetric tensor
fields.  This regularization scheme applies to all loop orders through a worldsheet conformal 
field formulated on a flat torus analytically continued to $6-\eps$ dimensions.  We shall show
shortly that all amplitudes are indeed IR regulated through this method, and that the naive
limit corresponds to the original amplitudes.

One may be worried that all self-dual, even tori cannot be properly formulated in
fractionalized dimensions.  However, the formulation of tori by background fields reveals that
this continuation is completely analogous to dimensional reduction and dimensional
continuation of ordinary field theory.  Also, the tori specified by lattices of signature
$(q+8n,q)$, with $n$ integer and $q>0$, are isomorphic to tori specified by lattices $\Gamma =
n\Gamma _8 \oplus q \Gamma _2$.  While the $E_8$ lattice cannot be deformed, the number $q$ of
two-dimensional lattices can be varied {\it continuously}.  Thus, there appears to be no
problem in analytically continuing in the number of lattice dimensions. 

Our regulator of choice is to multiply the internal modes by the lattice contribution from
$-2\eps$ scalars on the circle.  For our purposes we regulate the infra-red divergences in the
compactified four-dimensional model by introducing into the superstring a combination of
$2\eps$-scalars living on the line and
$-2\eps$ scalars on the circle, followed by ignoring the dependence on the vertex operators.
This choice maintains the central charge of the matter at $c=10$, thus no Weyl anomalies are
naively introduced. The regulator should be thought of only as that, a regulator that
maintains the symmetries of the amplitude, because
$-2\eps$ bosons on the world-sheet does not really define a pure model (in  the geometric sense
found by turning on background fields $G_{ij}$ and $B_{ij}$).  An alternative way of
regulating the IR would be to curve the target space-time, or  introducing addtional fields
but perhaps these regulators are too difficult to calculate in because of the possible 
introduction of the Liouville mode . 

The lattice contribution in eq.(\use\lattice), $Z(\Gamma)$, defined by the $g=1$ 
contribution for the lattice $\Gamma$, is 
$$ 
Z(\Gamma) = \sum_{P_L, P_R\in\Gamma} e^{i\pi \tau P_{\rm L} 
\c P_{\rm L} - i\pi \bar{\tau} P_{\rm R} \c P_{\rm R} } \ , 
\eqn\seclat 
$$ 
where the lattice momenta are given in general through (\use\latticemom).  
The expression in eq.(\use\seclat) may be regularized by subtracting $\eps$-bosons 
lying on the same or on a different lattice.  For the example of an orthogonal lattice (discussed
extensively  in [\use\gsb ]), we have the simple form for the lattice contribution 
from a single $U(1)$ factor, 
$$ 
Z_{\rm o} = (\tau _2 )^{{1 \over 2}}\sum_{m,n} e^{-2\pi mn 
\tau_1 - \pi \tau_2 ( m^2 a^2 + {n^2\over a^2} )} \ .
\anoneqn 
$$ 
However, as discussed previously, we may regularize by analytically continuing the lattice
metric, anti-symmetric tensor field, and gauge fields for the heterotic string to fractional
dimensions.  We shall continue to denote this regularization away from the 
integer dimensions by $Z_{\rm o}$; the case in which $Z_{\rm o}$ denotes 
an orthogonal lattice in $\eps$ dimensions being a special case of the construction.  

Putting the above ingredients together, we have the following defined
regularized partition function of the internal modes
$$ 
Z_{\rm reg}(\Gamma) = Z(\Gamma)~ Z_{\rm o} ^{-\eps} \ .
\anoneqn 
$$ 
It is important to note that whichever the regulator, the behavior for large $\tau_2$ is the
same, as can be seen from the following definition
$$  
Z_{\rm reg}(\Gamma) = (\tau_2 )^{{10-d\over 2}-\eps} {\tilde Z}_{\rm reg}(\Gamma)  
\anoneqn 
$$ 
Here, and elsewhere, the partition functions labelled with a tilde represent the lattice
sums without the $\tau _2$ prefactor.  For example, ${\tilde Z}(\Gamma)$ is given by the
lattice sum
$$ 
{\tilde Z}(\Gamma) = \sum_{m,n} |q|^{N_{m,n}} e^{2\pi i \tau_1 \phi _{m,n}} \ ,
\anoneqn 
$$ 
with $m$ and $n$ ranging over sets of integers that parametrize $P_L$ and $P_R$, and
$$ 
N_{m,n} = \half (P_L \c P_L + P_R \c P_R) \qquad \qquad 
\phi _{m,n} = \half (P_L \c P_L - P_R \c P_R) 
\eqn\lattnum 
$$ 
where the left and right lattice momenta are defined through eq.~(\use\latticemom).  
This discussion is easily extended to the case of the heterotic string, where 
the $Z$-factor is $z_i$-dependent.

\vskip .3in
\section{Analytic Structure} 
\vskip .2in 

The convergence properties of the amplitudes in eqs.~(\use\twobamp) and (\use\hetamp) are closely
related to those for the corresponding uncompactified one.  We  shall in this section concentrate
on the integral representations in eqs.~(\use\twobint)  and (\use\hetint).   The lattice
contributions are easily seen to not modify the exponential behavior in the large $\tau_2$ region
(each term gives additional factors of
$q$ and ${\bar q}$ which suppress the  large $\tau_2$ behavior).  Following the analysis of
[\use\dpone ], we see that absolute convergence holds only in the domain
$$  {\rm Re} ~s_{ij} = 0 \ ,
\anoneqn 
$$  for fixed non-zero $\eps >0$. Infrared convergence results from the  extra
$\eps$-dependence in the $\tau_2$ factors in $Z_{\rm reg}$ for $\eps >0$.

Given a convergent definition of the amplitude at the point Re($s_{ij})=0$, we now proceed  to
construct a unique analytic continuation that will extend to all momemta. We shall modify some
of the arguments given for the critical string in [\use\dpone ] for the dimensionally regularized 
version.  Also, we shall discuss only the Type {\rm IIb} string and only quote the final results for
the heterotic string.  

We begin by setting $z_4=0$ and separating the integration over the  three remaining locations
of the vertex operators into the regions, 
$$   0 \leq {\rm Im}~ z_1 \leq {\rm Im}~ z_2 \leq {\rm Im}~ z_3 \leq \tau_2  \ ,
\eqn\primord 
$$   and the permutations of $(1,2,3)$.  The different regions, labelled by  the orderings of
the three locations of the vertex operators, give rise  to a decomposition of the integral
representation in eq.(\use\twobint) into 
$$  A(s,t,u) = 2 A(s,t) + 2 A(t,u) + 2 A(u,s)  
\anoneqn 
$$  where $u = -s-t$ and $s= -2k_1\c k_2$, $t= -2k_2\c k_3$.  We  only need to work with
$A(s,t)$ since the remaining two subamplitudes are found by permutations of the momenta. 

With the ordering in (\use\primord) we introduce new variables $w_{ij}$  satisfying $\vert
w_{ij} \vert \leq 1$, defined by 
$$
\eqalign{ w_{ij} & = e^{2\pi i (z_i - z_j)} \quad\quad {\rm Im}~ z_{ij} > 0 \cr & = q e^{2\pi i
(z_i - z_j)} \quad\quad {\rm Im}~ z_{ij} < 0 }\ ,   
\anoneqn 
$$  with which we shall express the integral in (\use\twobint).   We shall also make use of
the standard parametrization of the vertex insertion  points in terms of the real variables
$\alpha _i$ and $u_i$
$$  z_i - z_{i-1} = {\alpha_i \over 2\pi} + i\tau_2 u_i, \quad\quad i=1,\ldots,4  \ ,
\anoneqn 
$$  where $\alpha _1 + \alpha _2 + \alpha _3 +\alpha _4 = 2 \pi \tau _1$  and $u_1 + u_2 +u_3
+u_4 =1$.  

In product expansion form, we may rewrite the prime form neglecting
the $z$ independent factors which will vanish as a result of momentum
conservation once we insert it into the expression for the four-point 
functions: 
$z_{ij}
\equiv z_i - z_j$,
$$
E(z_i,z_j)={\vartheta_1\bigl[z_{ij};\tau\bigr]\over\vartheta_1'\bigl[0;\tau\bigr]} 
  = \prod_{n=0}^{\infty} \bigl( 1- q^n e^{2\pi i z_{ji}} \bigr) 
 \bigl( 1- q^{n+1} e^{-2\pi i z_{ji}} \bigr) \ .
\anoneqn
$$ 
As usual we have defined $q\equiv e^{2\pi i \tau}$.  Next we define, 

$$
{\cal R}(w_{ij}) = \prod_{i<j} \vert {\rm E}(z_i, z_j) \vert^{2 k_i \c k_j}
 = \prod_{i\neq j} \prod_{n=0}^{\infty} \vert 1- w_{ij} q^n \vert^{-s_{ij}} \ .
\anoneqn
$$  
which is the component $e^{\half s_{ij} G(z_i,z_j)}$ of the amplitude, without 
including the zero mode substraction in the Greens function in eq.~(\use\twopoint).  

The amplitude $A(s,t)$ may then be rewritten as 
$$
\eqalign{  
A(s,t) = \int_{F} d^2\tau & {1\over \tau_2^{-1+\eps}}
\int^{2\pi} \prod^{4}_{i=1} {d\alpha_i \over 2\pi}  ~
  \delta (2\pi\tau_1 - \sum_j \alpha_j)
\cr &
\times\int^1 \prod_{i=1}^4 du_i \delta (1-\sum_j u_j) ~ 
\vert q \vert^{(-s u_1 u_3 - t u_2 u_4)} {\cal R}(w_{ij}) ~ {\tilde Z}(\Gamma) 
 {\tilde Z}_{\rm c}^{-2\eps} \ . } 
\anoneqn 
$$ 
The function ${\cal R}$ is defined from the product expansion of the $\vartheta$-functions
and may be expanded in an infinite series expansion as follows:
$$
\eqalign{ {\cal R}(w_{ij}) = & \prod_{i\neq j} \prod_{n=0}^{\infty} 
\vert 1- w_{ij} q^n \vert^{-s_{ij}}
\cr = & \prod _{i=1} ^4 \bigl | 1 - e^{i \alpha _i} |q|^{u_i} 
\bigr | ^{-s_i} \sum _{n_i=0} ^\infty \sum _{|\nu _i|\leq n_i} P^{(4)} _{\{n_i \nu _i\}} (s,t)
\prod _{i=1} ^4 |q|^{n_i u_i} e^{i \nu _i \alpha _i} \cr \ . } 
\eqn\breakup 
$$ 
Here, $s_i=s$ for $i$ even, $s_i=t$ for $i$ odd, and $P^{(4)} _{\{n_i \nu _i\}} (s,t) $ are  
polynomials in $s$ and $t$, which have previously been generated recursively 
in Appendix C of [\use\dpone ].

The explicit analytic continuation of the toroidal one-loop  amplitude for the four point
function and the singularities in the external momenta are then described by the following 
statement:
\bigskip
\noindent{\bf Toroidally Compactified Partial Amplitudes} 
\medskip

{\it For any integer N, we can break the original amplitude in eqs.(\use\twobamp) and 
(\use\hetamp) into the sum 
$$ 
A(s,t;\eps) = \sum_{\{n_i;m,n\}\in D_N} \sum_{\vert\nu_i\vert\leq n_i} 
P^{(4)}_{\{n_i,\nu_i\}}(s,t) A_{\{r_i,\eta_i\}}(s,t) + M_{N}(s,t;\eps)  \ ,
\eqn\firstthm 
$$ 
with
$$ 
r_i = n_i + N_{m,n} \qquad \eta_i = \nu_i + \phi_{m,n} \ .
\eqn\rfunctors 
$$ 
\smallskip\noindent
Here, $M_{N}(s,t)$ is a meromorphic function of $s$ and $t$ in the region of ${\rm
Re}(s)$ or
${\rm Re}(t) < N$, and the numbers $D_N$ range through $D_N=\{m,n;n_i ~~{\rm such~
that}~~  n_1+n_2+n_3+n_4 +N_{m,n}\leq 4N\}$. The  partial amplitudes
$A_{\{n_i,\nu_i\}}(s,t)$ are almost identical to  those encountered for the critical
superstring (together with the  Theorem 1 of [\use\dpone ] describing their analytic
behavior)  and are obtained by adjusting only the power of $\tau _2$ }: 
\smallskip 
$$
\eqalign{ A_{\{r_i,\eta_i\}} = \int_1^{\infty} &{d\tau_2 \over \tau_2^{-3+d/2+\eps}} 
\int^{2\pi}_0 \int_0^1 \prod^{4}_{i=1} {d\alpha_i du_i \over 2\pi}  ~ \delta (1 - \sum_j u_j ) ~
\vert q \vert^{(-s u_1 u_3 - t u_2 u_4)} \cr &
\times \prod_{j=1}^4 \bigl \vert 1 - e^{i\alpha_j} \vert q \vert^{u_j} 
\vert^{-s_{j,j+1}}
\vert q \bigr \vert^{r_i u_i} e^{i\eta_i \alpha_i} \ . } 
\eqn\partialamps 
$$ 
\bigskip 

The above reduces the analytic continuation of the amplitudes in eq.~(\use\twobint) and 
eq.~(\use\hetint) to the analytic continuation of the much simpler amplitudes $A_{r_i
\eta _i}$, where there are no longer any infinite products, and where all $\alpha_i$
integrations are decoupled. The proof is completely analogous to that of Theorem 1 in
[\use\dpone] and it will not be given here. In fact, the polynomials $P^{(4)}_{\{n_i \nu
_i\}} (s,t) $ defined through the  representation in eq.~(\use\breakup) are the same as
those that appear in the analysis of the four-point function (of states in the
gravitational multiplet)  of the critical string.  The only modifications from the
critical string are thus the modified power of $\tau_2$, due to the altered number of
uncompactified dimensions, and the shifts in the spectrum resulting from the addition of
$N_{m,n}$ and $\phi_{m,n}$ in the definition of $r_i$ and
$\eta_i$ arising from the compactification. These modifications are exactly the expected
ones from the  masses of the states in the toroidally compactified theory.  

The structure of the branch cuts and the poles on top of branch cuts of the four-point
amplitude arise in the individual amplitudes $A_{\{r_i,\eta_i\}}(s,t)$, which we now
examine.  

\vskip .2in
\section{Dispersion Relations} 
\vskip .2in 

In order to manifest the cut structure without an $i\epsilon$-prescription, 
we need to resort to old-fashioned dispersion theory.  
In this section we complete the analysis of the construction of the
dispersion relations necessary to describe the partial amplitudes 
$A_{\{r_i \eta_i\}}(s,t)$ in eq.~(\use\partialamps).  The analysis involves finding the 
dimensionally regularized versions of the spectral representation of the 
box diagram and its applications to study the analytic form of these partial 
amplitudes.  For ease of presentation, we break the analysis into three 
subsections.     

The crucial ingredient in studying the analytic form of the partial amplitudes, 
as found in [\use\dpone ], is the inverse Laplace transform $\varphi_{n \nu}$ of the
hypergeometric function.  This is defined as follows 

$$ 
\eqalign{ 
\int^{2\pi}_0 {d\alpha\over 2\pi} e^{i\alpha\eta} \vert 1 -  x e^{i\alpha} \vert^{-s} & =
C_{\vert\eta\vert}(s) x^{\vert\eta\vert}
 F({s\over 2}, {s\over 2}+ \vert\eta\vert; \vert\eta\vert + 1; x^2)  
\cr & 
= x^{-r}
\int^{\infty}_0 d\beta ~ x^{\beta} \varphi_{r\eta} (s;\beta) \ , }
\eqn\inverselap 
$$   where $\varphi_{r\eta} (s;\beta) = 0$ for $\beta < 0$. 

We first rewrite $A_{\{r_i,\eta_i\}}(s,t)$ using the inverse Laplace transform of the
hypergeometric function, 

$$
\eqalign{ A_{\{r_i,\eta_i\}}(s,t) = \int^{\infty}_0 \prod_{j=1}^4 & d\beta_j 
\varphi_{\{r_j, \eta_j\}}(s,t;\beta_i) ~\int^{\infty}_1  {d\tau_2 \over \tau_2^{-1 +\eps}}
\int^1_0 \prod_{i=1}^4 du_i \delta(1-\sum_j u_j)
\cr &
\times e^{2\pi\tau_2 (su_1 u_3 + tu_2 u_4 - \sum_{i=1}^4 u_i\beta_i)} \ .} 
\anoneqn 
$$ 
A number of changes of variables, identical to those used  in [\use\dpone ], leads to 

$$
\eqalign{ 
A_{\{r_i,\eta_i\}} = &
\int^{\infty}_0 \prod_{j=1}^4 d\beta_j \Psi_{\{r_j, \eta_j\}}  (s,t;\beta_i)~ \int_0^1 du_1
\int_0^{1-u_1} du_2 ~ (1-u_1-u_2) \cr &
\times \int_1^\infty d\mu \int_0^1 d\alpha {1\over \mu^{-1+\eps}}  
e^{-2\pi\mu\kappa} \ , 
} 
\eqn\partialamp 
$$ 
with the definitions  

$$
\kappa \equiv u_1\beta_1 + u_2\beta_2 + (1-u_1-u_2) 
\Bigl\{ \alpha (\beta_3 - su_1) + (1-\alpha) (\beta_4 - tu_2) \Bigr\} \ ,   
\anoneqn 
$$ 
and 
$$  
\Psi_{\{ r_j\eta_j\}}= \prod_{j=1}^4 \phi_{\{r_j\eta_j\}}(s;\beta_j) \ .
\anoneqn 
$$ 
The $\mu$-integral ranges from $0$ to $1$, 

$$
\int^1_0 {d\mu \over \mu^{-1+\eps}} e^{-2\pi\kappa\mu} =  E(\kappa) \ . 
\eqn\efunction 
$$  
Keeping the dimension $\eps<2$ gives $E(\kappa)$ an entire function in all its variables. Note
that this region would correspond to the ultraviolet portion of the box integral, which is
divergent for
$d\geq 8$.  Thus far we have obtained the analog in the toroidal   compactified case to Lemma 2
in [\use\dpone ], which is that the contribution of
$E(\kappa)$ to the partial amplitude is globally {\it meromorphic} in both
$s$ and $t$ and may also be absorbed into the meromorphic function $M_N (s,t;\eps)$.

Since we may absorb any meromorphic contribution from $A_{\{r_i,\eta_i\}}$ into the function
$M_N(s,t;\eps)$ we may extend the lower limit of the integration region of $\mu$ in
eq.(\use\partialamp) from one down to zero, at the cost of changing 
$M_N(s,t;\eps)$. Then we have after performing the $\mu$-integral explicitly, 

$$
\eqalign{ 
A_{\{r_i,\eta_i\}} = &
\int^{\infty}_0 \prod_{j=1}^4 d\beta_j \Psi_{\{r_j, \eta_j\}}  (s,t;\beta_i)~ \int_0^1 du_1
\int_0^{1-u_1} du_2 ~ (1-u_1-u_2) \cr &
\times \Gamma(2-\eps) \int_0^1 d\alpha (2\pi\kappa)^{-2+\eps} \ .} 
\eqn\almost 
$$ 
We see that the $\beta$-integral in eq.~(\use\almost) is over the regulated box 
integral functions $B(\beta_i;s,t;\eps)$ possessing masses $\beta_j$, defined by 

$$ 
B(\beta _i;s,t;\eps) = \Gamma(2-\eps) \int_0^1 du_1 \int_0^{1-u_1} du_2 ~ (1-u_1-u_2)
\int_0^1 d\alpha ~(2\pi\kappa)^{-2+\eps} \ .
\eqn\box 
$$ 
The box integral in eq.~(\use\box), although in a non-standard form, may be recast in the form of
a dispersion relation.  (Given  an $i\epsilon$ prescription, we may actually find the complete
expression  for the string amplitude because all of the dimensionally regularized box
integrals are known explicitly, at least in terms of dilogarthms [\use\bdkint ].)  

\vskip .2in 
\subsection{Dispersive Form of Regularized Box} 
\vskip .2in 

In this section we modify the integral form of the dimensionally regularized 
box diagram in (\use\box) into a form suitable for expressing the amplitudes 
in a dispersive form.  To find the box dispersion relation, we first perform 
the $\alpha$ integral in eq.~(\use\box), leading to 

$$ 
\eqalign{ 
B(\beta _i;s,t;\eps) & = 
\int^1_0 d\alpha (2\pi\kappa)^{-2+\eps} = - {\Gamma(1-\eps)\over \Gamma(2-\eps)} 
    \bigl[ 2\pi(1-u_1-u_2) \bigr]^{-2+\eps} 
\cr & 
\times{1\over y} 
\Bigl\{ (x_o+\beta_4 -tu_2 +y)^{-1+\eps} -  (x_o+\beta_4 -tu_2)^{-1+\eps} \Bigr\}  \ , }  
\anoneqn 
$$ 
where $y= \beta_3 -su_1 -\beta_4+tu_2$.  Then we use the identity 
\vskip .2in
$$
\int^{\infty}_0 dx {x^{\mu} \over (x+a) (x+b)} = {\Gamma(1+\mu)\Gamma(-\mu) 
\over b - a} \Bigl( b^\mu - a^\mu \Bigr) \ , 
\anoneqn 
$$ 
to rewrite the $\alpha$ integral into 

$$ 
\eqalign{ 
\Gamma(2-\eps)~\int^1_0 (2\pi\kappa)^{-2+\eps} = & {(2\pi)^{-2+\eps} \over 
\Gamma(\eps)} (1-u_1-u_2)^{-2+\eps} ~
\cr & 
\times \int_0^\infty dx {x^{-1+\eps}\over  (x+ x_o+\beta_3-su_1)(x+x_o+\beta_4-tu_2) } \ . } 
\anoneqn 
$$ 
We represent the two linear factors in the denominator by spectral integrals over 
the real variables $\sigma$ and $\tau$: 
$$ 
{1\over x+x_o+\beta_3-su_1} = \int_0^\infty d\sigma {1\over \sigma-s}  
   \delta(x+x_o+\beta_3-\sigma u_1) \ , 
\anoneqn 
$$  
and
$$
{1\over x+x_o+\beta_4-tu_2} = \int_0^\infty d\sigma {1\over \tau-t}  
   \delta(x+x_o+\beta_4-\tau u_2) \ .  
\anoneqn 
$$
Using these, we have the final representation of the regularized 
box graphs,  

$$   B(\beta_i;s,t;\eps) =
\int^\infty_o d\sigma \int^\infty_0 d\tau  {\rho_B(\beta _i;s,t;\sigma,\tau;\eps)\over
(\sigma-s)(\tau-t)}  \ ,
\anoneqn 
$$  
where the dimensionally regularized double spectral density $\rho_B$ is given by 

$$
\eqalign{
\rho _B(\beta _i;s,t;\sigma,\tau;\eps) = & {(2\pi)^{-2+\eps} \over \Gamma(\eps)} 
\int^1_0 du_1 \int^{1-u_1}_0 du_2 ~ (1-u_1-u_2)^{-1+\eps} \cr & \times \int^\infty_0 dx
~x^{-1+\eps} ~\delta(x+x_o+\beta_3-\sigma u_1) ~ 
\delta(x+x_o+\beta_4-\tau u_2) \cr} \ .
\eqn\almosttwo 
$$ 
The integrals in eq.~(\use\almosttwo) may be evaluated to give 

$$
\rho_B(\beta_i;s,t;\sigma,\tau ;\eps) = {(2\pi)^{-2+\eps} \over \Gamma(\eps)} 
~{1\over\sigma\tau}
\bigl({\sigma+\tau \over\sigma\tau}\bigr)^{-1+\eps}  (A^2-B^2)^{-1+\eps+{1\over 2}}
\vartheta(A^2-B^2) \vartheta(A) \ , 
\eqn\rhobeqn 
$$  where $A$ and $B$ are defined as,

$$  A = {\sigma\tau - (\beta_1+\beta_3)\tau - (\beta_2 + \beta_4)\sigma 
\over 2(\sigma+\tau)}
\qquad \qquad B^2 = {\beta_1 \beta_3 \tau + \beta_2 \beta_4 \sigma 
\over \sigma +\tau} \ . 
\anoneqn 
$$
The domain of support for the spectral density of the dimensionally regularized  box diagram with
masses
$\beta_j$ ($j=1,\ldots,4$) is found by solving the three constraints imposed by the theta
functions on
$A$, $A^2-B^2$ in eq.~(\use\rhobeqn) and from the fact that $\sigma,~ \tau \geq 0$.   

In the following we parameterize the solution to these three conditions 
and the allowed values of $\sigma,\tau$.   The first
constraint $A^2 - B^2$ becomes, after rewriting $B^2$ which involves an arbitrary mass scale
$m^2$, 

$$  
\eqalign{ 
\Bigl[ 1 - {\beta_1 + \beta_3 \over \sigma} -
 {\beta_2 + \beta_4 \over \tau} \Bigr]^2
& + \Bigl[ {\beta_1 \beta_3 - m^4 \over \sigma m^2} +
    {\beta_2 \beta_4 - m^4 \over \tau m^2} \Bigr]^2
\cr &  
\geq  \bigl[  {\beta_1 \beta_3 + m^4 \over \sigma m^2} +
    {\beta_2 \beta_4 + m^4 \over \tau m^2} \bigr]^2 \ . 
}
\eqn\firstconstraint
$$
Next we define the variables $x$ and $y$, 

$$
\Bigl[ {\beta_1 \beta_3 - m^4 \over \sigma m^2} +
 {\beta_2 \beta_4 - m^4 \over \tau m^2} \Bigr]
 = x \Bigl[  {\beta_1 \beta_3 + m^4 \over \sigma m^2} +
    {\beta_2 \beta_4 + m^4 \over \tau m^2}  \Bigr]
\anoneqn
$$
$$
\Bigl[ 1 - {\beta_1 + \beta_3 \over \sigma} -
 {\beta_2 + \beta_4 \over \tau} \Bigr]
= y \Bigl[  {\beta_1 \beta_3 + m^4 \over \sigma m^2} +
    {\beta_2 \beta_4 + m^4 \over \tau m^2} \Bigr] \,
\anoneqn
$$
so that eq.~(\use\firstconstraint) becomes the hyperbolic equation:

$$
x^2 + y^2 \geq 1 \ .
\anoneqn
$$
The next constraint $\sigma, \tau\geq 0$ then becomes a bound for
$x$, so that its allowed range is $x_{-} \leq x \leq x_{+}$ with

$$
x_{\pm} = \bigl[ {\beta_1 \beta_3 - m^4 \over \beta_1 \beta_3 + m^4} ,
~ {\beta_2 \beta_4 - m^4 \over \beta_2 \beta_4 + m^4} \bigr] \ .
\eqn\secondconstraint
$$
The last constraint $A\geq 0$ becomes the statement $y\geq 0$,
since the inequality has the form,

$$
A = {\sigma\tau \over 2(\sigma + \tau)} \bigl( {\beta_1 \beta_3 + m^4 \over
 \sigma m^2} + {\beta_2 \beta_4 + m^4 \over \tau m^2 }\bigr) y \ .
\anoneqn
$$
The parameterization of the region above in $\sigma$ and $\tau$ in
terms of $x$ and $y$ and the masses $\beta_j$ has the solution to $\sigma$ and
$\tau$:

$$
{1\over\sigma} = {A_2 \over \lambda} \quad\quad {1\over\tau}
 = - {A_1 \over\lambda} \ ,
\anoneqn
$$
with 

$$
\eqalign{
A_1 & = \beta_1 \beta_3 - m^4 - x(\beta_1 \beta_3 + m^4)
\cr
A_2 & = \beta_2 \beta_4 - m^4 - x(\beta_2 \beta_4 + m^4)
\cr
\lambda & = (\beta_1 + \beta_3) A_2 - (\beta_2 + \beta_4) A_1 - 2yDm^2
\cr
D & = \beta_1 \beta_3 - \beta_2 \beta_4 \ .
}
\anoneqn
$$
For a given mass configuration $\beta_i$, the spectral
density vanishes if $\sigma$ and $\tau$ are outside the region specified
by the allowed space of $x$ and $y$ above:  The minimum values for
$\sigma$ and $\tau$ are found to be,

$$
(\beta_1 + \beta_3)^2 \quad\quad (\beta_2+\beta_4)^2 \ .
\eqn\branchvalues
$$
The box diagram in eq.~(\use\box) possesses cuts at values of $s$ and $t$ 
beginning at the values in eq.(\use\branchvalues), which is the expected 
two-particle threshold condition. 

\vskip .2in 
\subsection{Dispersive Form of Partial Amplitudes} 
\vskip .2in 

For the string partial amplitude in eq.(\use\partialamps) the inverse Laplace transform of the
hypergeometric function leads to the factors $\Psi_{\{r_i \eta_i\}}$, which can be interpreted as
an infinite sum of Dirac point masses:  

$$
\Phi_{r_i \eta_i} = \sum_{k_i =0} C_{k_i}(s_i) C_{k_i + \vert \eta_i \vert}(s_i) 
\delta(2k_i+r_i+\vert \eta_i \vert - \beta_i) \ ,
\eqn\diracpoints 
$$ 
and 
$$
\Psi_{\{r_i \eta_i\}} = \prod_{j=1}^4 \Phi_{r_j \eta_j} \ . 
\anoneqn  
$$   
The functions $C_{k_i}(s_i)$ are defined through the inverse Laplace  transform in
eq.~(\use\inverselap).  The total spectral density can be factorized into, after
performing the $\beta$ integration, the following form: 

$$
\rho(s,t;\sigma,\tau;\eps) = \sum_{k_i = 0}^{\infty} \prod_{j=1}^4  
  C_{k_j}(s_j) C_{k_j + \vert \eta_j \vert} (s_j) ~ 
\rho_B(\beta_j; \sigma,\tau;s,t;\eps) \vert \ .
\eqn\finalrho 
$$ 
The line, $\vert$, in eq.~(\use\finalrho) denotes that the $\beta_i$ 
are equal to $\beta_i= 2k_i+r_i+\vert \eta_i \vert$.

Combining the above results, we arrive at the following theorem: 
\vskip .2in 
\noindent
{\bf Dispersive Form of Toroidally Compactified Partial Amplitudes} 
\medskip 
\noindent
{\it The partial amplitudes $A_{r_i \eta _i} (s,t)$  can be expressed as 

$$ 
A_{\{r_i \eta _i\}} (s,t) = M_{\{ r_i \eta _i \}}  (s,t) + \int _0 ^\infty d\sigma \int _0
^\infty d \tau {\rho _{\{ r_i \eta _i\}} (s,t;\sigma,\tau) \over (s-\sigma) (t-\tau ) } \ ,
\eqn\disppartamp 
$$ 
where the double spectral density is given by 

$$
\rho_{\{ r_i \eta _i\}} (s,t) = \int_0 ^\infty \prod _{j=1} ^4  
  d \beta_j ~\varphi_{\{r_j \eta_j\}} (s,t;\beta_j) \rho_B  
  (\beta_j;s,t;\sigma,\tau;\eps) \ .
\eqn\secondthm 
$$  
It is assumed that we maintain the number of uncompactified dimensions $d<8$.  In this 
case no ultra-violet divergences arise within the box graph:  We do not need to 
perform any subtractions in the double dispersion representation.  The double 
spectral density in eq.~(\use\secondthm) has been regularized and leads 
to an infra-red regularized expression for the partial amplitudes 
$A_{\{r_i \eta _i\}} (s,t)$.} 
\vskip .2in

The term $M_{\{r_i \eta _i\}}(s,t)$ in eq.~(\use\disppartamp) is a globally 
meromorphic function of $s$ and $t$, similar to the 
contribution $M_N(s,t)$ in eq.~(\use\firstthm); for
example, it receives contributions from changing the lower limit of integration in
eq.~(\use\partialamp), producing the meromorphic result in eq.~(\use\efunction).  

We may denote the contribution of each box diagram that arises from the 
integral over the spectral density in eq.~(\use\secondthm) by the notation 
$A_{\{r_i \eta _i\}}^{m_i^2} (s,t)$.  The internal masses $m_i^2$ on the 
lines of the box depend on the value of the particular $\beta_i$ pulled out 
by the integration over $\sigma$ and $\tau$ in eq.~(\use\disppartamp).  The 
support for the box diagrams has been found above, and the branch cuts in 
$s$ and $t$ appearing from the integration in eq.~(\use\disppartamp) begin 
at $\beta_i=0$; further ones appear at the discrete values of $\beta_j$ 
extracted from the integration over the Dirac point masses in eq.~(\use\diracpoints).  

The condition on the masses $\beta_j=0$ correspond to the
minimum mass values, 

$$  
m_j^2\vert_{\rm min} =  r_j + \vert\eta_j\vert \ .
\eqn\minmass
$$  
\smallskip\noindent
Thus the integral representation on the right hand side of eq.~(\use\disppartamp) 
defines a holomorphic expression in $s$ and $t$ in the cut plane
$s,~t\in {\bf C} - {\bf R}_+$: 

$$ 
(s,t) \in \Bigl( ~{\bf C} \backslash [ (M_1^2 + M_3^2)^2, \infty ) ~\Bigr) 
\times \Bigl( ~{\bf C} \backslash [ (M_2^2 + M_4^2)^2, \infty ) ~\Bigr) \ . 
\anoneqn 
$$  
\smallskip\noindent 
with the mass parameters $M_i^2 = 2k_i+r_i +\vert\eta_i\vert$.  The lattice 
momenta contribute to the locations of the branch cuts through $N_{m,n}$, 
defined in eq.~(\use\lattnum),  

$$ 
N_{m,n} = \half (P_L \c P_L + P_R \c P_R) \qquad  
 \phi_{m,n} = \half (P_L \c P_L - P_R \c P_R) \ ,
\anoneqn 
$$  
and enters through the definitions $r_i=n_i+N_{m,n}$ and $\eta_i=\nu_i+\phi_{m,n}$ 
respectively, from eq.~(\use\rfunctors).

There are several features of the dispersive form of the amplitude worth 
pointing out.   First, as already noted
above  we have isolated the locations of the origins of the branch cuts in 
the values of $(s,t)$; these points define the masses for particle 
production in intermediate string states.  Second, the one-loop amplitude 
is genuinely unitary in the double dispersive representation, and may be used 
for example to check against the optical theorem.  Last, we may use the 
integral representation to construct an effective $i\epsilon$-prescription, 
as in [\use\dpone ].


\vskip .3in 
\section{Discussion} 

In this work we have extended the construction initially used to define the four-point 
function in critical string theory to the cases of a toroidal compactification scheme 
in both the IIb and Heterotic models.  The pole structure and unitarity of the 
compactified string amplitudes are found through  defining them via single and 
double dispersion relation.  Furthermore, we have provided an regularization 
scheme which is used to isolate the infrared singularities in the amplitudes.  The 
scattering amplitudes we
consider are those of the scattering of states in the Neveu-Schwarz/Neveu-Schwarz sector;
however, more general amplitudes at four-point are certainly amenable to these techniques.  

In the future it would be quite interesting to explore the application of these methods to
amplitudes containing external fermion creation operators, in which case possible ambiguities
in the integration over supermoduli space might appear [\use\ars ].  Additionally, the
construction should also be applied to the more general case of higher-point amplitudes in
both the critical and non-critical case.  However, in the former case the spectral
representation of the amplitudes does not admit a unique analytic continuation through the
construction given here [\use\cdp ].  In part II in of this work we shall
present the  analogous construction for amplitudes within orbifold compactification 
schemes [\use\parttwo ].    

\vskip .3in 
\noindent { \bf Acknowledgements} 
\vskip .2in 

The author acknowledges Eric D'Hoker and D.H. Phong for many useful discussions and 
collaboration on this work.  This research was supported  the by National Science
Foundation grant PHY-97-22101 and PHY-92-18990.  


\vfill\break 

\listrefs

\end